\documentclass[prl,twocolumn]{revtex4}

\usepackage{epsfig,graphicx}
\usepackage{amsmath,amsfonts,amssymb}

\newcommand{\unit}{\leavevmode\hbox{\small1\kern-3.6pt\normalsize1}}

\newcommand{\be}{\begin{equation}}

\def    \bea           	{\begin{eqnarray}}
\def    \eea           	{\end{eqnarray}}

\newcommand{\Mp}{M_P}

\newcommand{\gravitino}{{\widetilde{G}}}
\newcommand{\mgravitino}{{m_{\widetilde{G}}}}

\newcommand{\photino}{\widetilde{\gamma}}
\newcommand{\zino}{\widetilde{Z}}
\newcommand{\wino}{\widetilde{W}}

\newcommand{\ra}{{\rightarrow}}

\begin{document}
\title{New  decay modes of gravitino dark matter}
\author{Ki-Young Choi}
\email{kiyoung.choi@pusan.ac.kr}
\affiliation{Department of Physics, Pusan National University, Busan 609-735, Korea}
\author{Carlos E. Yaguna}
\email{carlos.yaguna@uam.es}
\affiliation{Departamento de Fisica Teorica and Instituto de Fisica Teorica UAM-CSIC\\
Universidad Autonoma de Madrid, Cantoblanco, E-28049 Madrid, Spain}

\begin{abstract}
We consider the three-body decays of gravitino dark matter in supersymmetric scenarios with bilinear $R$-parity violation. In particular, gravitino decays into $\ell W^*$ ($\ell f\bar f'$) and $\nu Z^*$ ($\nu f\bar f$) are examined for gravitino masses below $M_W$. After computing the gravitino decay rates into these three-body final states and studying their dependence on supersymmetric parameters, we find that these new decay modes are often more important than the two-body decay, into a photon and a neutrino, considered in previous works. Consequently, the gravitino lifetime and its branching ratios are substantially modified, with important implications for the indirect detection of gravitino dark matter. 
\end{abstract}

\maketitle

\section{Introduction}
Recent measurements indicate that about $25\%$ of the energy density of the Universe is made up of a mysterious form of non-baryonic matter known as  cold dark matter \cite{Dunkley:2008ie}. To explain it, physics beyond the  Standard Model is required. Supersymmetric extensions of the Standard Model are the most promising scenarios that may account for  the dark matter. They are motivated, in addition, by the hierarchy problem and by the unification of the gauge couplings.

Within supersymmetric models, R-parity conservation was long believed to be a
prerequisite for supersymmetric dark matter, for it guarantees the  stability
of the lightest supersymmetric particle (LSP) --the would-be dark matter
candidate. It was pointed out in \cite{Takayama:2000uz}, however, that if  the gravitino --the superpartner of the graviton that  arises in local
supersymmetric theories-- is the LSP, it  can be a suitable dark matter candidate  in R-parity violating models.
 Indeed, even though the gravitino is unstable in such scenarios, 
 its lifetime may be much longer than the age of the Universe. Such a long lifetime is the result of the gravitino feeble interactions, which are suppressed by the Planck scale and by  small R-parity violating couplings. A scenario with gravitino dark matter and R-parity violation is even favoured by  thermal leptogenesis, as it alleviates the tension between the large reheating temperatures required by leptogenesis and the constraints from  Big-Bang Nucleosynthesis \cite{Buchmuller:2007ui}. In R-parity violating models, therefore, the primordial gravitino produced in the early Universe is  a  viable and well-motivated dark matter candidate.

An important feature of these R-parity violating models is that, unlike their R-parity conserving counterparts, gravitino dark matter can be indirectly detected. In fact, the small fraction of gravitinos that have decayed until today constitutes a source of high energy cosmic rays \cite{Bertone:2007aw,Ishiwata:2008cu}. In their decays, gravitinos may produce gamma rays \cite{Bertone:2007aw,Ibarra:2007wg}, neutrinos \cite{Covi:2008jy} and antimatter \cite{Ibarra:2008qg}  that could be observed in present and future experiments. Present data from FERMI, for instance, already constrains the parameter space of these models \cite{Choi:2009ng}.

The indirect detection signatures of gravitino dark matter  strongly depend on the gravitino lifetime and on its branching ratios. In previous works, it has been assumed that they are determined by the two-body decays of the gravitino. We will show, in this paper, that that is not always the case. For gravitino masses below $M_W$, the three-body final states $\ell W^*$ and $\nu Z^*$, where * denotes a virtual particle, typically give a large contribution to the gravitino decay width, modifying in a considerable way the gravitino lifetime and its branching ratios. These results have important implications for scenarios of decaying gravitino dark matter, especially regarding their indirect detection prospects.
\section{Two-body decays}
We work in the framework of  supersymmetric models with bilinear R-parity
violation with gravitino as LSP, such as those considered in~\cite{Buchmuller:2007ui,Ibarra:2007wg,Ishiwata:2008cu}. In these models, the sneutrino field acquires a vacuum expectation value  that breaks lepton number and gives rise to mixing between neutralinos and neutrinos and between charginos and charged leptons. Gravitinos may then decay into standard model particles via lepton number violating interactions.

Because the R-parity violating couplings are expected to be largest for the third generation, we will assume, following~\cite{Covi:2008jy}, that the sneutrino acquires a vev only along the $\tilde\nu_\tau$ direction. It is convenient to define a dimensionless R-parity breaking parameter as   $\xi_\tau=\langle\tilde\nu_\tau\rangle/v$, where $v= 174$ GeV. This parameter is constrained from neutrino masses and from successful BBN to be in the range \cite{Ishiwata:2008cu}
\begin{equation}
10^{-11}\lesssim \xi_\tau \lesssim 10^{-7}.
\end{equation}
In this setup, all R-parity violating processes, including the decay of the  gravitino, are controlled by $\xi_\tau$.

Gravitinos can decay into two-body final states containing either a neutrino
(and a photon, $Z$ boson, or higgs boson) or a charged lepton (and a $W$
boson). Because R-parity is broken along the $\tilde\nu_\tau$ direction,  the final state charged
lepton is  a $\tau$  and the neutrino is a $\nu_\tau$. The gravitino decay
rates can be calculated directly from the interaction Lagrangian~\cite{Bagger:1990qh}. For the two-body final states $\gamma\nu_\tau$, $Z\nu_\tau$ and $W^+\tau^-$, they  are 
given by~\cite{Ishiwata:2008cu,Covi:2008jy}

\begin{align}
\Gamma(\gravitino\ra \gamma \nu_\tau)&=
\frac{\xi_\tau^2\mgravitino^3}{64\pi\Mp^2}|U_{\photino\zino}|^2,\\
\Gamma(\gravitino\ra Z \nu_\tau)&=
\frac{\xi_\tau^2\mgravitino^3}{64\pi\Mp^2}\beta_Z^2 p_Z(\mgravitino,M_Z,U_{\zino\zino}),\\
\Gamma(\gravitino\ra W^+ \tau^-)&=
\frac{\xi_\tau^2\mgravitino^3}{32\pi\Mp^2}\beta_W^2 p_W(\mgravitino,M_W,U_{\wino\wino}) ,\label{eq:2bodyW}
\end{align}
where
\begin{equation}
p_Z=\left[|U_{\zino \zino}|^2
  f_Z - \frac83\frac{m_Z}{\mgravitino} Re[U_{\zino \zino}]j_Z+\frac16h_Z \right]
\end{equation}
and $\beta,f,j,h$ are simple functions --see \cite{Covi:2008jy} for their definition.  In this expression,  $U_{\photino\zino},U_{\zino\zino},U_{\wino\wino}$ are, respectively, the photino-zino, the zino-zino, and the wino-wino mixing parameters \cite{Covi:2008jy}. To a good approximation they can be written as \cite{Ibarra:2007wg,Grefe:2008zz}
\begin{align}
|U_{\photino\zino}|&\approx \frac{M_Z(M_2-M_1)s_Wc_W}{(M_1c_w^2+M_2s_w^2)(M_1s_W^2+M_2c_W^2)},\\
|U_{\zino\zino}|&\approx \frac{M_Z}{M_1s_W^2+M_2c_W^2},\\
|U_{\wino\wino}|&\approx \frac{M_W}{M_2},
\end{align}
where $s_W$ and $c_W$ denote the sine and cosine of the weak mixing angle, and
$M_1,M_2$ are the $U(1)$ and $SU(2)$ gaugino masses. For simplicity, we will
assume that gaugino masses are universal at the GUT scale, so that $M_2\sim
1.9 M_1$ at the electroweak scale. Notice, from the above equations, that the mixing parameters decrease with gaugino masses.

The two-body decays of gravitino dark matter have been studied in detail in previous works~\cite{Ishiwata:2008cu,Covi:2008jy,Buchmuller:2009xv}. For gravitino masses below $M_W$, gravitinos decay into $\gamma\nu_\tau$ with a $100\%$ branching ratio. Above the $W$ threshold, this decay mode becomes quickly negligible, accounting for less than $1\%$ of the branching for gravitino masses larger than $150$ GeV. If heavier than the $W$ boson, the gravitino  dominantly decays into $W^\pm\tau^\mp$, with significant contributions from $Z\nu_\tau$ (once it is open) and, in some cases, from $h\nu_\tau$. The main point of this paper is that to properly compute the gravitino lifetime and its branching ratios the two-body decays considered so far in the literature are not enough; a certain class of three-body decays must also be taken into account.

\section{Three-body Decays}
\begin{figure}[t]
\begin{center} 
\begin{tabular}{ccc}
\includegraphics[scale=0.25]{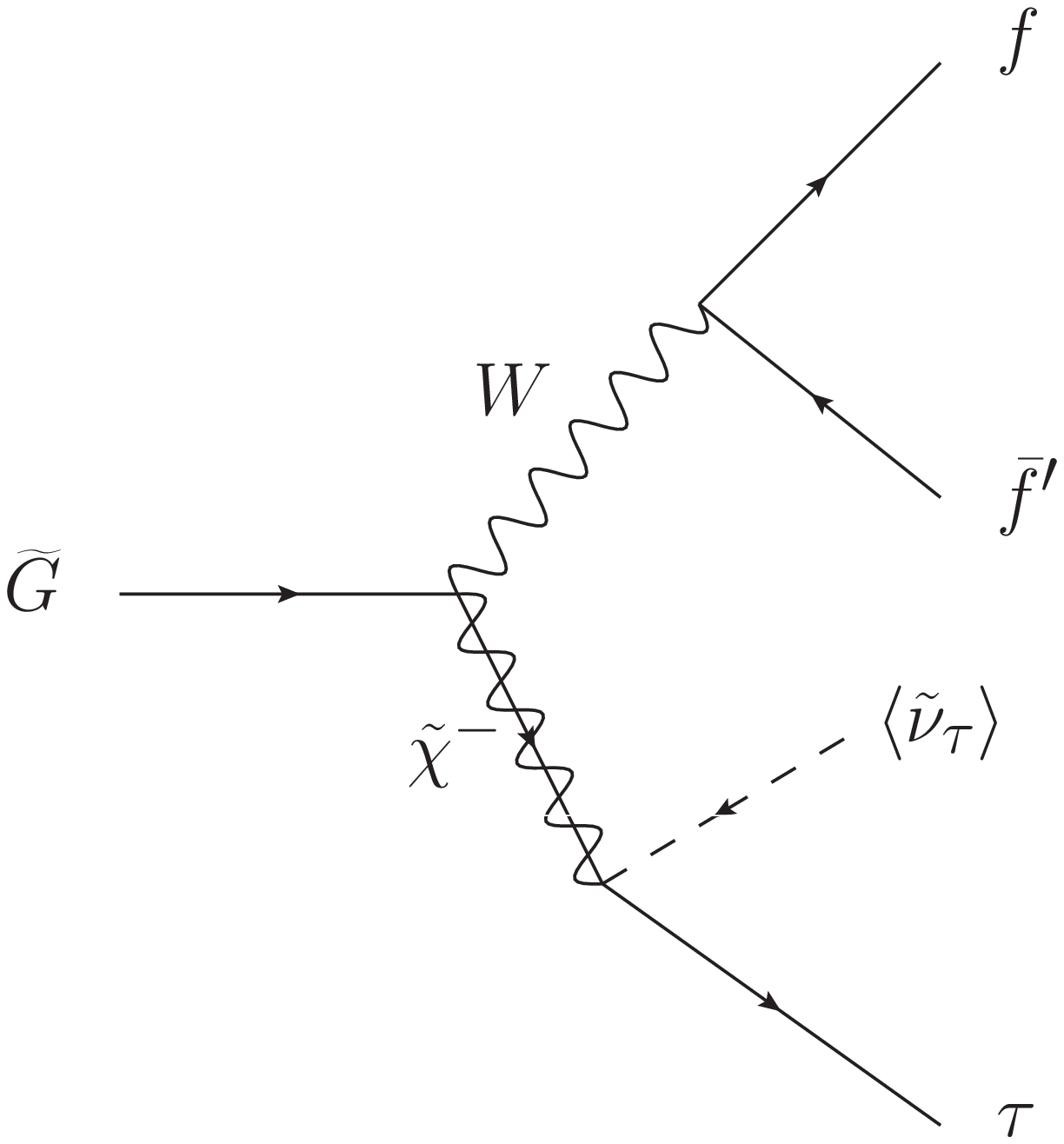} & \hspace{1cm} &\includegraphics[scale=0.25]{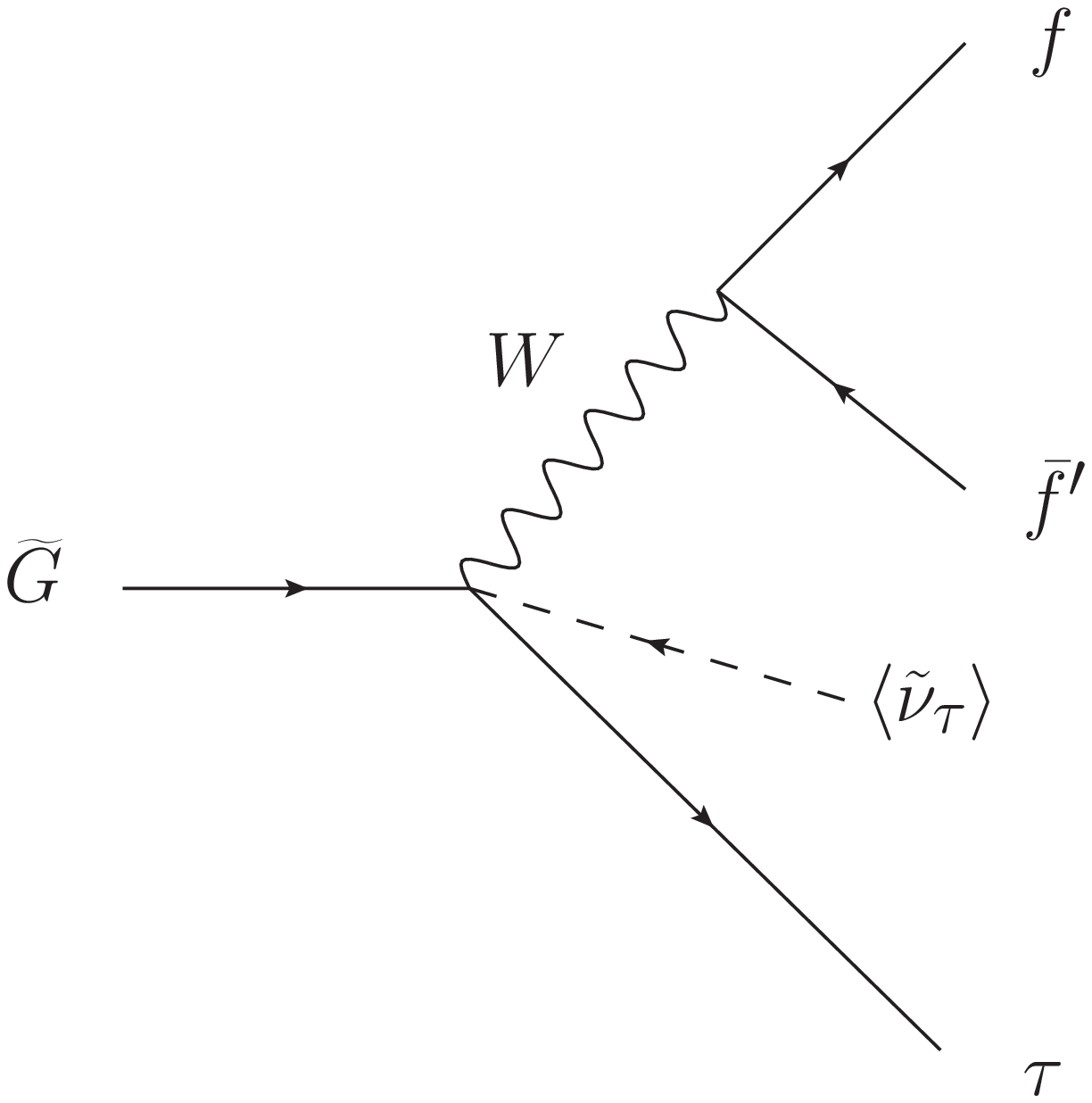}\\
(a) & \hspace{1cm} & (b)
\end{tabular}
\caption{The feynman diagrams contributing to the gravitino decay into $\tau W^*$.\label{fig:diagW}}
\end{center}
\end{figure}
 For gravitino masses smaller than $M_W$ the only two-body channel kinematically available for the decay of the gravitino is $\gamma \nu_\tau$. In addition to it,  there exists three-body final states consisting of a lepton and a virtual massive gauge boson, such as $\tau W^*$ ($\tau 
f\bar f'$) and $\nu_\tau Z^*$ ($\nu_\tau f\bar f$), that may contribute significantly to the gravitino decay width. These three-body final states are considered for the first time in this paper.

Two diagrams contribute to the decay of the gravitino into the three-body
final state $\tau W^*$ ($\tau f \bar f'$) -- see figure \ref{fig:diagW}. The contribution of the (a) diagram is proportional to $U_{\wino\wino}$, whereas (b), which comes from the non-abelian $4$-vertex,  is independent of it.  We have computed, from these two diagrams, the decay rate $\Gamma(\gravitino\to \tau^-+W^{+*}\to \tau^-f\bar f')$.  The differential decay rate  can be written as
\begin{equation}
\frac{d\Gamma}{dsdt}=\frac{N_c}{256\pi^3\mgravitino^3}\overline{|{\mathcal M}|^2},
\end{equation}
where $N_c$ is the color factor, and $s$ and $t$ are, respectively, the invariant masses for the $f\bar f'$ and the $f\tau$ systems. The squared amplitude can be factorized as
\begin{equation}
\overline{|{\mathcal M}|^2}=   \frac{g^2\xi_\tau^2}{64\Mp^2}\times f(\mgravitino,U_{\wino\wino},m_f,s,t),
\label{eq:msq}
\end{equation}
where $f$ is a long function of the given parameters. As usual, the decay rate is obtained from the differential one by integrating over the possible values of $s$ and $t$.

An important feature of the function $f$ is that, due to the interference between the diagrams $(a)$ and $(b)$ in figure \ref{fig:diagW}, it contains terms quadratic in, linear in, and independent of $U_{\wino\wino}$.  These latter terms, which are analogous to those found in equation (\ref{eq:2bodyW}), will play a crucial role in our analysis, as they  favour the decay into the three-body final states for larger gaugino masses. 

The gravitino decay into $\nu_\tau Z^*$ proceeds through  diagrams analogous
to those from figure \ref{fig:diagW}, and the expression for its decay rate
has a similar form. In principle, the decay into $\nu_\tau\gamma^*$  should
also be taken into account, as it interferes with the $\nu_\tau Z^*$
contribution. We have verified, however, that the photon-mediated diagram  is
very much suppressed and can be neglected. In bilinear R-parity breaking models, other three-body final states are not expected to be relevant for gravitino decays.

The total decay width of a gravitino with $\mgravitino < M_W$ will, then, be given by
\begin{equation}
\frac{\Gamma_{tot}(\gravitino)}{2}= \Gamma(\gravitino\to \gamma\nu_\tau)+\Gamma(\gravitino\to \tau^+ W^{-*})+ \Gamma(\gravitino\to \nu_\tau Z^*),
\end{equation}
where the factor two takes into account the charge conjugated final states. This decay width is  proportional to the R-parity breaking parameter $\xi_\tau^2$, see equation (\ref{eq:msq}).  The gravitino lifetime, $\tau_\gravitino$, is simply given by the inverse of the decay width, $1/\Gamma_{tot}(\gravitino)$. In the following we study the gravitino lifetime and branching ratios as a function of  supersymmetric parameters. 
\begin{figure}[t]
\begin{center} 
\includegraphics[scale=0.3]{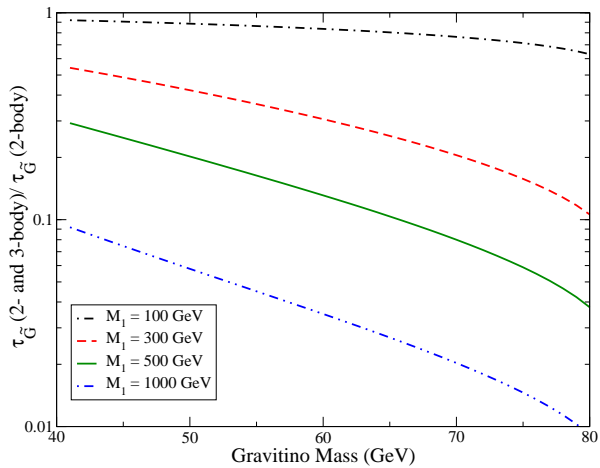}
\caption{The ratio between the gravitino lifetime computed from two- and three-body decays and that one obtained from the two-body final state only as a function of the gravitino mass for different values of $M_1$.\label{fig:gdecay2lts}}
\end{center}
\end{figure}

To begin with, let us consider the effect of the three-body final states,
$\tau W^*$ and $\nu_\tau Z^*$, on the gravitino lifetime. Figure
\ref{fig:gdecay2lts} displays, as a function of the gravitino mass, the ratio
between the correct gravitino lifetime (computed from two- and three-body
final states) and the gravitino lifetime obtained from the two-body final
state only. This ratio, which is independent of $\xi_\tau$, would be equal to $1$ if the contribution from three-body final states were negligible. Clearly, that is not the case. The  lines in that figure  correspond to different values of the gaugino masses.  As expected, the heavier the gauginos, the larger the effect of the three-body final state. Notice that there is a significant deviation from the two body result over a wide range of gravitino and gaugino masses. For $M_1=300$ GeV, for instance, the effect of the three-body final states is significant already for gravitino masses around $40$ GeV, yielding a lifetime about a factor two smaller than that computed for two-body final states; as the gravitino mass increases the effect becomes larger, reaching an order of magnitude close to $M_W$. For $M_1=1$ TeV, the genuine gravitino lifetime can be more than two orders of magnitude smaller than that obtained from two-body decays. Undoubtedly, the three-body final states $\tau W^*$ and $\nu Z^*$ have a considerable impact on the computation of the gravitino lifetime.
\begin{figure}[t]
\begin{center} 
\includegraphics[scale=0.3]{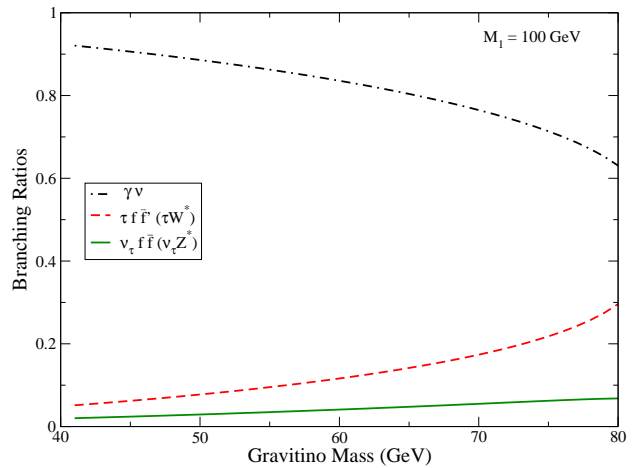}
\caption{The gravitino branching ratios as a function of the gravitino mass for  $M_1=100$ GeV.\label{fig:gdecay2brs100}}
\end{center}
\end{figure}

Besides the lifetime, the other important quantities that determine the indirect detection signatures of gravitino dark matter are its decay branching ratios. Figure \ref{fig:gdecay2brs100} compares the two-body and three-body decay branching ratios of the gravitino as a function of $\mgravitino$ for $M_1=100$ GeV. For this value of $M_1$ the dominant decay mode of the gravitino is the two-body final state $\gamma\nu_\tau$, but with a  branching ratio that can be much smaller than $1$, reaching about $60\%$ for $\mgravitino\sim M_W$.  We also notice from the figure that, among the two three-body decay final states, it is the $\tau W^*$ that dominates. Its branching can reach $30\%$ for $\mgravitino\sim M_W$. Remarkably, even for small gaugino masses, the contribution of the three-body final states is significant.

\begin{figure}[t]
\begin{center} 
\includegraphics[scale=0.3]{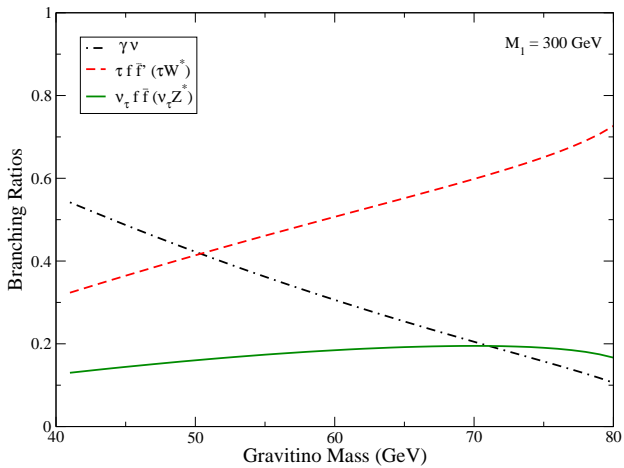}
\caption{The gravitino branching ratios as a function of the gravitino mass for  $M_1=300$ GeV.\label{fig:gdecay2brs}}
\end{center}
\end{figure}

For slightly larger gaugino masses, the branching ratios look quite different, as illustrated by figure \ref{fig:gdecay2brs}. This figure is analogous to the previous one but for $M_1=300$ GeV. Notice that, in this case, the two body final state dominates only for gravitino masses below $50$ GeV. From that point up to $M_W$, it is the $\tau W^*$ final state that accounts for most gravitino decays. Close to $M_W$, the $\tau W^*$ final state has a branching of about $70\%$. It is also observed in the figure that the $\nu Z^*$ branching can be quite significant, reaching  $20\%$  and  becoming even more important that the two-body decay for gravitino masses between $70$ GeV and $M_W$. This figure demonstrates that three-body final states may easily dominate the gravitino decays.

\begin{figure}[t]
\begin{center} 
\includegraphics[scale=0.3]{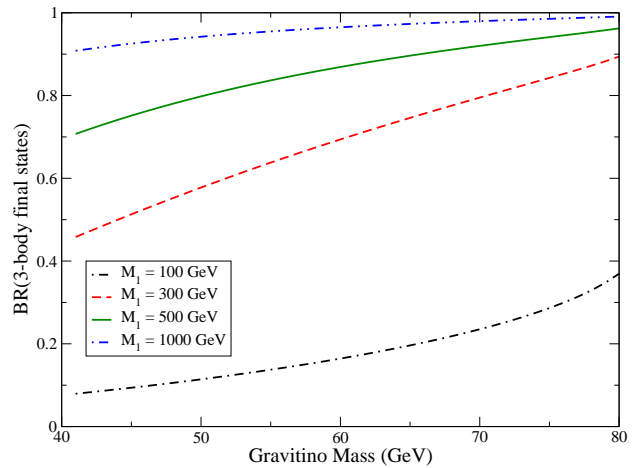}
\caption{The gravitino branching ratio into three-body final states as a function of the gravitino mass for different values of $M_1$.\label{fig:gdecay2}}
\end{center}
\end{figure}

To better study its dependence with gaugino masses, we show, in figure \ref{fig:gdecay2}, the branching ratio into three-body final states, $BR(\gravitino\to \tau W^*)+BR(\gravitino\to \nu_\tau Z^*)$, as a function of the gravitino mass for different values of $M_1$. Notice that the three-body final states are always significant and that they dominate the gravitino decays even for moderate values of the gaugino masses. If $M_1=1$ TeV, the three-body final states have a branching larger than $90\%$ over the entire gravitino mass range we consider, $40\mathrm{GeV}<\mgravitino<M_W$.

In view of these new results on the gravitino lifetime and its branching ratios, existing constraints on scenarios with decaying gravitino dark matter have to be reexamined, and the prospects for the  indirect detection of gravitino dark matter have to revised. Because of the smaller branching into $\gamma\nu_\tau$, the gamma ray line will be less pronounced than previously believed, and it will be accompanied by a  continuum of gamma rays coming from  the three-body decays of the gravitino. Besides, the possibility of discovering gravitino dark matter via antimatter opens up for gravitino masses below $M_W$, thanks to the positrons and antiprotons generated by the $\tau W^*$ and $\nu_\tau Z^ *$ final states. In a future work, we will study these interesting topics in detail.
\section{Conclusions}
We have studied the decays into $\tau W^*$ and $\nu_\tau Z^*$ of gravitino dark matter in the context of supersymmetric models with bilinear R-parity violation. We computed the gravitino decay rates and showed that these previously neglected processes typically  give large contributions to the gravitino decay width. They actually dominate the decay rate  over a wide range of gravitino and gaugino masses. In a future work, we will study in detail the implications of these results for gravitino dark matter.
\begin{acknowledgments}
K.Y. Choi was partly supported by the Korea Research
Foundation Grant funded by the Korean Government (KRF-2008-341-C00008)
and by the second stage of Brain Korea 21 Project in 2006.
C. E. Y. is supported by the \emph{Juan de la Cierva} program of the MICINN of Spain, by  the CAM under grant HEPHACOS S2009/ESP-1473 and by the MICINN Consolider-Ingenio 2010 Programme under grant MULTIDARK CSD2009-00064.
\end{acknowledgments}

\end{document}